\centerline{\bf HINTS OF A NEW RELATIVITY PRINCIPLE FROM }
\centerline{\bf FROM $p$-BRANE QUANTUM MECHANICS}
\bigskip
\centerline{ Carlos Castro}
\centerline{Center for Theoretical Studies of Physical Systems} 
\centerline{ Clark Atlanta University, Atlanta, GA. 30314, USA}
\bigskip
\centerline{ December 1999} 
\bigskip

\centerline { \bf ABSTRACT}
This report is an extension of previous one hep-th/9812189. Several quantum mechanical wave equations for $p$-branes are proposed. 
The most relevant $p$-brane quantum mechanical wave equations determine the quantum dynamics involving the creation/destruction of $p$-dimensional loops of topology $S^p$,  
moving in a $D$ dimensional spacetime  background, in the quantum state $\Phi$. To implement full covariance we are forced to enlarge the ordinary Relativity principle to a 
$new$ Relativity principle, suggested earlier by the author based on the construction of {\bf C}-space, and also  by Pezzaglia's Polydimensional Relativity, 
where all dimensions and signatures of spacetime should be  included on the same footing.

\bigskip

\centerline{\bf I. {INTRODUCTION}}
\bigskip

Quantization of $p$-branes is a notoriously difficult unsolved problem. This work a modest step towards the eventual solution of such problem and raises the 
possiblity of a new Relativity principle operating in Nature which may reveal important clues behind the geometrical origins of $M$ theory. 
We present here an extension of previous work [19] where several different quantum mechanical wave equations associated with the quantum dynamics of $p$-branes based on 
two  views are furnished : (i)  that $p$-branes can be seen as gauge theories of volume preserving diffeomorhisms and (ii) that closed $p$-branes of topology $S^p$ may be 
represented by $p$-dim loops moving in spacetime.  

One of these wave equations represents the quantum field theory dynamics of $p$-branes living in a $D$ dimensional spacetime with a given 
world volume measure configurations ( Jacobians)  in a quantum state $\Psi$. The other wave equations determine the quantum dynamics involving the creation/destruction  
of $p$-dimensional loops of topology $S^p$ in a $D$ 
dimensional spacetime  background ,  in the quantum state $\Phi$. A square root  procedure a la Dirac yields wave equations of first order. 
The last wave equation based on the 
Kanatchikov-Navarro De Donder-Weyl quantization program is a step towards a finite dimensional formulation of the field theory of $p$-branes.

Comparisons with previous loop equations are made [5] . To implement full covariance,  we are forced to enlarge the ordinary Relativity principle to a 
$new$ Relativity principle ,  suggested earlier by the author [8] based on the construction of {\bf C}-space, and also  by Pezagglia's Polydimensional Relativity [9] , 
where all dimensions and signatures of spacetime should be  included on the same footing. 

The outline of this work goes as follows : In section {\bf 2.1 } we discuss the $p$-loop wave equations and in {\bf 2.2} the Duality between $p$-branes and Composite Antisymmetric Tensor Field Theories [4] . 
In {\bf 3.1 } we compare other loop wave equations and discuss the need for a New Relativity principle. Some preliminary results on the New Relativistic tranformations in 
{\bf C}-space are given in {\bf 3.2} . In {\bf 4.1 } the Dirac equation for $p$-branes  is  presented and in {\bf 4.2}  the     
Kanatchikov-Navarro De Donder-Weyl quantization program towards a finite dimensional formulation of the field theory of $p$-branes is finally given.

\bigskip
\centerline{\bf 2. p-Brane Wave Equations suggest a New Relativity Principle }
\bigskip

\centerline{\bf 2.1 Wave Equations based on Antisymmetric Tensor Field Theories}
\bigskip
A local antisymmetric tensor gauge field theory reformulation of extended objects was given by [1,2] which later paved the way for the author to show that $p$-branes could be seen as composite antisymmetric tensor gauge field theories of the volume-preserving diffeomorphisms group [3,4] associated with the target  space of the primitive scalar field constituents . The action [1,2] was defined  :

$$S=-g^2 \int d^D x \sqrt { - {1\over (p+1)!}
(W_{\nu_1 \nu_2....\nu_{p+1}})^2 } +      
{1\over (p+1)!}W^{\nu_1 \nu_2....\nu_{p+1}} \partial_{\nu_1} B_{ \nu_2....\nu_{p+1}}. \eqno (2.1)$$
$ B_{ \nu_2....\nu_{p+1}}$ was a Lagrange multipier enforcing the transversality condition on the antisymmetric current $ W^{\nu_1 \nu_2....\nu_{p+1}} $. 
The field equations associated with the $W$ and $B$ fields admit $p$-brane solutions when the antisymmetric tensor current $W$ evaluated on any spacetime point $x^\mu$, is related to the $p+1$-vector tangent to the worlvolume of the $p$-brane  :

$$ W_{\nu_1 \nu_2....\nu_{p+1}}(x) =\kappa \int d^{p+1}\sigma 
\delta^D (x -X(\sigma )) {\vec X}_{\nu_1 \nu_2....\nu_{p+1}} (\sigma). \eqno (2.2)$$
with the $p+1$-vector tangent to the $p$ brane's worldvolume  :

$$ {\vec X}_{\nu_1 \nu_2....\nu_{p+1}} (\sigma)=\epsilon^{\sigma^1 \sigma^2....\sigma^{p+1}}\partial_{\sigma^1}X_{\nu_1}......\partial_{\sigma^{p+1}}X_{\nu_{p+1}}=
{\partial (    X_{\nu_1}, X_{\nu_2},.....X_{\nu_{p+1}}  )\over \partial (\sigma^1,\sigma^2,......,\sigma^{p+1}           ) }. \eqno (2.3)$$
Plugging   the solution (2.2, 2.3) into the original action (2.1) yields the $p$-brane Dirac-Nambu-Goto action after using the property of the delta function integration :

$$S=-(\kappa g^2) \int d^{p+1}\sigma \sqrt { - {1\over (p+1)!}
({\vec X}_{\nu_1 \nu_2....\nu_{p+1}})^2 }. ~~~T_p=\kappa g^2. \eqno (2.4)$$
the $p$-brane tension is expressed in terms of the coupling constant $g$ with the addition of an extra parameter $\kappa$ to accomodate dimensions : $ T_p=\kappa g^2$. The solution (2.2) is defined up to a scaling parameter given by $\kappa$.  
The polymomenta ${\Pi}_{\nu_1 \nu_2....\nu_{p+1}}$ variables conjugate to the $integrated$  $p+1$-tangent hyperplanes to the world volume of the $p$-brane are defined as :

$${\Pi}_{\nu_1 \nu_2....\nu_{p+1}} \equiv {\partial S \over \partial {\vec X}^{\nu_1 \nu_2....\nu_{p+1}} }= 
T_p {  {\vec X}_{\nu_1 \nu_2....\nu_{p+1}}\over  \sqrt { - {1\over (p+1)!}
({\vec X}_{\nu_1 \nu_2....\nu_{p+1}})^2 } } \eqno (2.5a)$$
We find it useful to remark the following technical subtlety, contary to the literature , that the polymomenta are  strictly speaking $not$ the conjugate variables to the 
Jacobians ${\vec X}_{\nu_1 \nu_2....\nu_{p+1}}$ given by  eq-(2-3) but to the $integrals$  of these Jacobians.  This parallels the fact that $p$ is the conjugate variable to the coordinate $q$ and 
$not$ to the velocity ${dq\over dt}$ in ordinary point particle mechanics. This technical issue will become clear below when we write down the wave functional equations 
associated with the quantum dynamics of $p$-dim loops of spherical topolgy, $S^p$, where each $p$-dim loop  is  seen as the $only$ boundary of an $open$ $p+1$-dim hypersurface.   

The $p$-brane extension of the Hamilton-Jacobi equation is [1,2] :

$$- {1\over (p+1)!} ({\Pi}_{\nu_1 \nu_2....\nu_{p+1}})^2 +T^2_p = 
- {1\over (p+1)!} [ {\partial (S^1(x), S^2(x),...,S^{p+1})\over      
\partial (x^{\nu^1}, x^{ \nu^2},....x^{\nu^{p+1}} ) } ]^2 +T^2_p =0. \eqno (2.5b)$$
where $S^1(x), S^2(x),...S^{p+1}(x)$ are the Clebsch Potentials [1,2]. The polymomentum ${\Pi}_{\nu_1 \nu_2....\nu_{p+1}}$, when extended from the $p$-brane's world tube to all of spacetime, is the so-called slope or geodesic field of the $p$-brane.  The pullback from spacetime to the $p$-brane's world tube of the slope/geodesic field is the $p$-brane's volume-momentum, 
$\Pi^{a^1}_{\nu_1} (\sigma^a)$    ,
conjugate  to the $p$-brane's configuration coordinate $X^\nu (\sigma^a), ~a=1,2,...p+1$ :

$${\Pi}_{\nu_1 \nu_2....\nu_{p+1}} \epsilon^{a^1 a^2....a^{p+1} }
\partial_{a^2}X^{\nu_2}(\sigma^a) \partial_{a^3}X^{\nu_3}(\sigma^a)......... 
\partial_{a^{p+1}} X^{\nu_{p+1}} (\sigma^a) \equiv \Pi^{a^1}_{\nu_1} (\sigma^a). \eqno (2.6)$$
 
The reformulation of $p$-branes in terms of antisymmetric tensor fields was based on the old vortex/hydrodynamics ideas of Nielsen and Olesen.    
$p$-branes are reinterpreted as higher dimensional vortices. The vortex velocity (inside the fluid) at any given point of the vortex matches the fluid's 
velocity at that given point. The vortex velocity plays the role of the $p+1$-vector tangent to the $p$-brane's world tube whereas the fluid's velocity plays the role of the slope/geodesic field, ${\Pi}_{\nu_1 \nu_2....\nu_{p+1}}$ defined outside the $p$-brane. When the slope/geodesic field is restricted to have support on the $p$-brane's world tube then it matches the $p+1$-vector tangent to the $p$-brane.         

We propose our first quantum mechanical wave functional equation for the closed $p$-brane, of topology $S^p$, a $p$-dim loop $C_p$,  based on the Hamilton-Jacobi equation to be of the following  form  :

$$ { 1 \over (p+1)!} { \delta^2 \Phi (C_p)  \over  \delta {\Omega }_{\nu_1 \nu_2....\nu_{p+1}}(C_p)  \delta {\Omega}^{\nu_1 \nu_2....\nu_{p+1}}(C_p)  } +
{{\cal E}^2_p  }\Phi =0. \eqno (2.7a)$$
which is just the constraint implementation of the Hamilton-Jacobi
on-shell condition (2.5)  on $\Phi (C_p) $. The explicit derivation of (2.7a) will be presented below. 
To make contact with the previous functional wave equations in the literature it proves convenient to write (2-7a) in the form :

$$ \oint d^{p}\mu (s)  ~[{ 1 \over (p+1)!} { \delta^2  \over  \delta {\Omega }_{\nu_1 \nu_2....\nu_{p+1}}(C_p)  \delta {\Omega }^{\nu_1 \nu_2....\nu_{p+1}}(C_p)  } +
   {\cal E}^2_p]\Phi (C_p)  =0. \eqno (2.7b)$$
the quantities in (2-7a) are function-valued objects whereas upon integration in (2-7b) they become ordinary numbers. $d^p\mu (s)$ is a $p$-dimensional reparametrization invariant measure on the boundary $\partial M = C_p$, a $p$-dim loop of topology $S^p$ parametrized by $(s^1,s^2,....s^p)$. . 

If one adopts the view taken by the literature that the polymomentum is the conjugate to the Jacobians (2-3), generalized velocities, then the Hamilton-Jacobi equation reads :

$$ \int d^{p+1}\sigma \sqrt {det~h}~  ~[{ 1 \over (p+1)!} { \delta^2  \over  \delta {\vec X }_{\nu_1 \nu_2....\nu_{p+1}}(\sigma )  \delta 
{\vec X}^{\nu_1 \nu_2....\nu_{p+1}}(\sigma )  } +
   { {\cal E}^2_p \over m^{2 (p+1)}}   ]\Psi  (   {\vec X }_{\nu_1 \nu_2....\nu_{p+1}} )  =0. \eqno (2.7c)$$
Some important remarks are in order. 

Eq-(2-7c) is  by construction only $p+1$-dimensional volume-preserving diffs invariant since the Jacobians (2-3) are volume-preserving diffs invariant by definition. 
One could fully covariantized (2.7c) by adding the judicious terms in the derivative operators. For the time being we shall not be concerned with this covariantization program. However, (2-7b) is explicitly $p$-dim boundary reparametrization invariant. 
 The fields $\Phi, \Psi$ are  taken to be a  Lorentz scalars since the Hamilton-Jacobi equation and the wave equations (2.7) must remain 
Lorentz invariant. In the next section we will take the square root of (2.7) a la Dirac to obtain a first order wave equation where the fields 
$\Phi, \Psi$ are  no longer scalars  nor trivial spinors as it occurs for a point particle. Similar studies were undertaken by Hosotani and collaborators [13-15 ]. The transformation laws of the fields under Lorentz transformations for the $p$-brane field theory are highly nontrivial since they involved the introduction of generalized Dirac 
matrices. 

We wish to emphasize that there are many differences between this work and that of [13-15]. Mainly their wavefunctional is defined solely in terms of the coordinates $X^\mu (\sigma)$ as $\Psi [X^\mu (\sigma)]$. 
These authors fix the temporal reparametrizations by choosing the gauge $X_0=\tau$  ; they work at the end in a simplified minisuperspace where their wave $functional$ equations are reduced yielding ordinary differential equations of a 
multicomponent field $\Psi$ depending solely on a finite number of variables, instead of an infinite number of components.  These variables are related to the locations and shapes of the $p$-branes : center of mass coordinates, length, volumes,.... and total ( integrated ) mean curvatures of the $p$-brane.  Their bosonic string spectrum does not contain tachyons in variance with the other quantization approaches. Nevertheless, the simplified wave equations furnished by Hosotani et al [13-15] will be very useful to find nontrivial solutions to the wave equations (2-7) other than the standard plane wave types. It is also desirable to find a relationship between our approach and theirs.    
 
${\cal E}_p$ is the energy per
unit volume of the $p$-brane or $p$-brane's tension. We set our units to be such $\hbar =1$ and $m$ in (2-7c) is a suitable parameter of mass needed to match dimensions . This is $not$ an artificial introduction of an extra parameter, $m$, but it has a well defined physical meaning in the Eguch-Schild quantization scheme of the string [5]. The usual Nambu-Goto-Polyakov quantization is based on keeping the string tension $constant$ while taking the average over $all$ worldsheet areas. The string tension is related to the Planck scale/Regge slope $\alpha '$  :  
$$T_2=m^2=               {1\over 2\pi \alpha '}. \eqno (2.8)$$ 
Whereas in the Eguchi-Schild quantization one starts from an action
given by the squared of the string's worldsheet area ( instead of the
area) and invariant under area-preserving diffs : 

$$ S= {1\over 2\pi \alpha' }\int d^2\sigma \{ X^\mu, X^\nu \}_{PB}\{ X_\mu, X_\nu \}_{PB}         \eqno (2.9)$$
the Poisson brackets are taken w.r.t the $\sigma^1,\sigma^2$ variables. 
One quantizes by keeping $fixed$ the world sheet area of the string histories 
and taking the average over all the $variable$ string tension values; i.e the energy per unit length of the string, ${\cal E}$ , is not kept constant. The Nambu-Goto quantization imposes the condition : ${\cal E}=T_2=m^2$ which is not the case in the Eguchi-Schild quantization scheme. 

Therefore, $m^{2(p+1)}$ is not necessary equal to ${\cal E}^2_p$.  
For these reasons we shall set the ratio $({\cal E}_p/m^{p+1})^2 $ to be equal to 
$\lambda^2$; i.e the ``eigenvalue'' of the functional derivative operator acting on $\Psi$. There is another version of the Schild action for strings and $p$-branes that is explicitly reparametrization invariant by means of introducing an auxiliary field . Upon elimination of this field via its equations of motion one recovers the Dirac-Nambu-Goto action. Such action is of the area, volume squared form with a cosmological constant terms added.  For the time being we shall not be concerned with this particular action. 
The quantity : 

$$\Psi=\Psi [ {\vec X}^{\nu_1 \nu_2....\nu_{p+1}} (\sigma^1,....,\sigma^{p+1})]. \eqno (2.10)$$  
is the wave $functional$ associated to a $p$-brane with a given $p+1$-vector tangent to the $p$-brane. 

${\vec X}^{\nu_1 \nu_2....\nu_{p+1}} (\sigma^a)$ are just the $components$ of the Jacobians (2-3) associated with the 
$p+1$ world volume of the $p$-brane as it sweeps spacetime ( a measure configuration)   in the quantum state $\Psi$. 
We must emphasize that this wavefunctional must not be interpreted as a probability amplitude. We are dealing with a truly covariant field theory of $p$-branes, which upon 
( second) quantization, becomes a quantum field theory of $p$-branes and $not$ a first quantized one. We will discuss shortly the quantization of the field theory of $p$-branes associated with the equation (2-7).

$\Psi$ truly represents a field which after second quantization will create/destroy a $p$-brane with a given tangent hyperplane  to the $p+1$ world volume at each point ( with a given measure configuration at each point ); i.e the    
$p$-brane's history embedded in  a $D$-dimensional target spacetime background in the quantum state $\Psi$ has a given tangent plane at each  point. Eq-(2-7c) is linked to the fact that $p$-branes are seen here as theories related to gauge theories of the volume-preserving diffs group. The latter are not ordinary non-abelian YM gauge theores that we are familiar with but composite antisymmetic tensor field theories [3]. The center of mass motion of the $p$-brane is factored out in these theories. For example, in the membrane case, this gives the 
$SU(N=\infty)$ gauge theory instead of the $U(N)$ one. Roughly speaking, the wave functional equations (2-7c) do not govern the center of mass motion ( the location) , only the volume configurations ( shapes)  of the $p$-brane.   

We present now the derivation of (2-7a,2-7b). We concentrate for simplicity on closed $p$-branes, a $p$-dimensional $loop$ , of topology $S^p$, whose history is  a $p+1$-dimensional  world volume embedded in a target spacetime of dimensions $D$. Such loop history emerges initially from the ``vacuum''  that is represented by a point : the 
$p$-loop starts intially  as a point and evolves in time ( the proper world volume time traced out by the $p$-dim loop in its motion through spacetime) to a given spatial configuration $C_p$ of topology $S^p$. At any given moment the $p$-loop can be seen as the only boundary of the $p+1$ world volume hypersurface spanned by the $p$-dim loop.  Based on this picture one has that an integration by parts yields :    

$$\Omega^{\mu_1\mu_2.....\mu_{p+1}}= 
\int_M  d^{p+1}\sigma  {\vec X} ^{\mu_1\mu_2.....\mu_{p+1}}
=\int_{\partial M} d^p s  X^{\mu_1}\wedge dX^{\mu_2}\wedge ....
\wedge d X^{\mu_{p+1}}. \eqno (2.11)$$ 
$\Omega^{\mu_1 \mu_2....\mu_{p+1}}$  are the $p+1$-dimensional volume ( holographic) components enclosed by the boundary, 
$C_p =\partial M $ ( a $p$-dimensional loop)   of the hypersurface ( world volume) $M$. 
The coordinates of the boundary or $p$-loop, $X^\mu (\vec s)$,  can be parametrized by the $p$ components $s=(s^1,s^2,....s^p)$ 
or also by the $\sigma$ variables subject to the conditions defining the boundary : $X^\mu (\sigma)| _{\partial M} = X^\mu (s^1,s^2,....s^p)$.  Such volume components  $\Omega^{\mu_1\mu_2.....\mu_{p+1}}$, or Plucker coordinates,  are those obtained inequivalently as the projections ( holographic shadows) of the $p$-loop onto the coordinate planes of the embedding space time. For the time being we shall not be concerned with the Plucker conditions which ensure us that we have a $1-1$ correspondence between the $p$-loop and the Plucker coordinates [5] . Neither we will be concerned at the moment with the different topologies assigned to the hypersurface. 
If there is no boundary then the integral (2-11) is automatically $zero$. One may envision this boundaryless case when the initial $p$-loop starts  from a point and evolves to a point : a vacuum to vacuum transition spanning a closed $p+1$-hypersurface : a $virtual$ $p$-brane history.   

In the case of the string, $p=1$, (2-11) becomes  :

$$\sigma^{\mu \nu}  [C_1] =\int dX^\mu \wedge d X^\nu ={1\over 2} \oint_{C_1}  (X^\mu dX^\nu -X^\nu dX^\mu). \eqno (2.12)$$
which is nothing but Stokes law. $\sigma^{\mu \nu}  [C_1]$ are the so-called holographic ( or shadows) area components of the loop $C_1$ projected onto the coordinates planes of spacetime [5]. .

Our view which differs from the literature is that the  $p+1$-volume components $ \Omega^{\mu_1\mu_2.....\mu_{p+1}}$ are the true conjugate variables to the polymomenta  of the closed $p$-brane 
$\Pi^{\mu_1\mu_2.....\mu_{p+1}}$ since the former is given by the integral of the latter :

$$\int d^{p+1} \sigma \sqrt { {1 \over (p+1)!} ||{\vec X} ^{\mu_1\mu_2....\mu_{p+1}}||^2 }~  
\Pi^{\mu_1\mu_2.....\mu_{p+1}} =T_p \int_M  d^{p+1}\sigma  
{\vec X} ^{\mu_1\mu_2.....\mu_{p+1}}=T_p \Omega^{\mu_1\mu_2.....\mu_{p+1}}. \eqno (2.13)$$
where we have used for the integration measure the quantity : 

$$\int d^{p+1} \sigma \sqrt {   {1 \over (p+1)!}     ||{\vec X}^{\mu_1\mu_2....\mu_{p+1}} ||^2 } =
\int d^{p+1}  \sigma \sqrt { det ~ h }   . \eqno (2.14)$$
This is the multi-vector ( multisymplectic) generalization of the canonical momentum of a point particle . For example, in nonrelativistic mechanics one has the ordinary momentum ${\vec p} =m {d{\vec X} \over d t}$.  
An integration of ${{\vec p} \over m} $ w.r.t the variable $t$ yields the coordinate 
${\vec X}$ which is the canonically conjugate variable to ${\vec p}$. The velocity is $not$ the canonical conjugate to the momentum. For this simple reason we refrain from saying that the Jacobians (2-3) ( generalized velocities) are the canonically conjugates to the  $\Pi^{\mu_1\mu_2.....\mu_{p+1}}$.

The bracket is then defined on the $p$-dim $boundary$  $\partial M$  : 
$$\{\Pi^{\mu_1\mu_2.....\mu_{p+1}} (s), \Omega_{\nu_1\nu_2.....\nu_{p+1} } (s') \}|_{\partial M}  =
\delta^{\mu_1\mu_2.....\mu_{p+1}}_{\nu_1\nu_2....\nu_{p+1}} \delta^{p} (s-s'). \eqno (2.15)$$
Upon quantization one has " equal time " commutators of the polymomenta and volume operators. The temporal variable in this picture is nothing but the invariant proper world volume interval spanned by the evolution of the 
$p$-dim loop given across its motion through spacetime and is given by the Dirac-Nambu-Goto action (2-11). This is the higher dimensional extension of the Eguchi $area$ quantization of the Schild string [5]. The areas, volumes,....are interpreted as the evolution $parameters$. It in this respect that the bracket operations are defined on the 
boundary; i.e on the $p$-dim loop $C_p$. Equal areal, volume,....time canonical quantization implies sitting on the $p$-dim loop at any given instant.  It is warranted to find a rigorous multisymplectic geometric formulation of the brackets (2-15) and the whole Hamilton-Jacobi equations (2-7, 2-17).

The usual prescription :

$$\Pi^{\mu_1\mu_2.....\mu_{p+1}} \rightarrow i {\delta\over \delta \Omega ^{\nu_1 \nu_2....\nu_{p+1}} (C_p) }. ~~~ (\hbar =1) \eqno (2.16)$$ 
renders the  Hamilton-Hacobi equation as a constraint on $\Phi (C_p) $ : 

$$ { 1 \over (p+1)!} { \delta^2 \Phi \over  \delta {\Omega}_{\nu_1 \nu_2....\nu_{p+1}}(C_p) \delta {\Omega}^{\nu_1 \nu_2....\nu_{p+1}} (C_p)  } +
{\cal E}^2_p  \Phi =0. \eqno (2.17)$$
To make contact with the literature we shall again write (2-17a) in the integral form : 

$$ \oint_{C_p} d^p \mu (s) ~[ { 1 \over (p+1)!} { \delta^2  \over  \delta {\Omega}_{\nu_1 \nu_2....\nu_{p+1}}(C_p) \delta {\Omega}^{\nu_1 \nu_2....\nu_{p+1}} (C_p)  } +
{\cal E}^2_p ]  ~ \Phi (C_p)=0. \eqno (2.18)$$
with $d^p \mu(s)$ being a $p$-dimensional reparametrization invariant measure of the $p$-dim boundary. Notice that having started with the Jacobian variables (2-3) that were invariant soley under volume-preserving diffs, eq-(2-17b) is however explicitly $p$-dimensional reparametrization invariant ; i.e invariant under $(s^1,s^2,....s^p)$ go to $({\tilde s}^1, {\tilde s}^2,....)$. The Lorentz scalar $\Phi =\Phi [{\Omega }^{\mu_1\mu_2....\mu_{p+1}} (C_p)] $ is the classical field functional of the $p$-dim loop $C_p$.

Upon ( second ) quantization the $\Phi$ becomes an operator which creates ( destroys)  the $p$-dim loops $C_p$. The functional derivatives appearing in (2-17) are $p$-loop functional derivatives that can be visualized as 
attaching an infinitesimal $p$-dim $petal$ to the $p$-loop at a given point along the $p$-loop. A $petal$ that encloses an infinitesimal volume whose holographic projections onto the coordinate embedding spacetime planes are what one means by $\delta {\Omega}_{\nu_1 \nu_2....\nu_{p+1}}(C_p)$.

If one adopts the conventional view of the literature one arrives at eq-(2-7c). Since a given $p$-loop is seen as the only boundary of the $p+1$ world volume ( hypersurface) associated with the history of a closed $p$-brane which began as a point ( loop shrinking to a point) and evolved to the spatial configuration $C_p$ tracing out an open $p+1$ world volume in a given time, one     
can associate to a given $p$-loop field $\Phi [\Omega] $ the corresponding $open$-hypersurface field $\Psi$   :

$$\Phi [{\Omega }^{\mu_1\mu_2....\mu_{p+1}} (C_p)] \leftrightarrow  \Psi 
[{\vec X }^{\mu_1\mu_2....\mu_{p+1}}(\sigma) ]. \eqno (2.19)$$
that allows to rewrite the initial Hamilton-Jacobi constraint equation on $\Phi$ also in terms of $\Psi$ like we did in eq-(2-7c). The latter is the field ( after second quantization) which 
creates ( destroys)  a given $p+1$-dim hypersurface with $C_p$ as its $only$  boundary and whose tangent plane  ( measure density ) configuration
at any given point of the hypersurface is  
${\vec X }^{\nu_1 \nu_2....\nu_{p+1}}( \sigma) $.

A String representation of quantum Loops has been given in [8] where the wavefunctional of a relativistic loop was explicitly expressed in terms of string degrees of freedom 
starting from a covariant $phase~ space$  Schild action path integral. The Polyakov string partition function was furnished from first principles and something more was obtained : the Eguchi wavefunctional, encoding the area dynamics of the boundary,  was also derived and, in addition, extra terms to regularize boundary ultraviolet divergences were included as well in one single scoop. A phase space path integral process for $p$-branes, similar to the one employed in [8] should render the specific relationship between $\Phi [C_p]$ and 
$\Psi [{\vec X }^{\nu_1 \nu_2....\nu_{p+1}}( \sigma)]$. Although this is a very difficult problem.     

\bigskip
\centerline{\bf 2.2 Branes/Composite Antisymmetric Tensor Field Theories : A New Duality Principle } 
\bigskip
Before ending this section we wish to bring an important connection between $p$-branes and nonabelian composite antisymmetric tensor field theories which are $not$ of the Yang-Mills type [4] . 
In [3] we have shown that $p$-branes can been seen as extended solutions to composite antisymmetric tensor gauge field theories of the volume preserving diffs group. Although 
the Lagrangians were of the YM types, which precisely bear the same resemblance as the Schild's Lagrangian  associated with the group $SU(\infty)$ of area-preserving diffs for spacetime independent gauge field configurations ( vacuum), these nonabelian theories are $not$ Yang-Mills theories. The Lagrangians [3] were :

$${1\over g^2_p}[{F}_{\nu_1 \nu_2....\nu_{p+1}} (\phi^a (x)) ]^2 . \eqno (2.20)$$
with the $composite$ field strength defined in [4] :

$${F}_{\nu_1 \nu_2....\nu_{p+1}} (\phi (x)) = \epsilon _{a_1 a_2......a_{p+1}}
\partial_{\nu_1} \phi^{a_1}(x) \partial_{\nu_2} \phi^{a_2}(x)....
\partial_{\nu_{p+1}}\phi^{a_{p+1}}(x). \eqno (2.21)$$
the fields $\phi^a (x)$ are a collection of $a=1,2,....p+1$ primitive scalars
living in a spacetime of dimensionality  $D=p+1+p'+1$ and taking values in a target space 
of dimensionality $p+1$. The $dual$ primitive scalars ${\tilde \phi}^b(x)$  
live in spacetime and take values in a target space of dimensionality $p'+1$.
Notice that this is the $dual$ picture to what we are familiar with  in the description of $p$-branes. Here we have maps from spacetime into manifolds of $less$ dimension : an immersion. We have shown  that  every strongly coupled composite antisymmetric tensor field theory of $p+1$-rank defined in $D$ spacetime has a weakly coupled $p'$ brane solution and vice versa. Also we have $T$ duality 
with a large/small volume duality in the space of solutions . $S$ and $T$ dualities were interrelated to each other. 
The relevance of these composite theories is that we have a formulation of $p$-branes where $S, T$ dualities are already 
built in from the very start : at the Lagrangian level, there is no need to conjecture them. 
 
There was another sort of duality [3,19] ] ( this duality has been recently discussed in the literature [20] by Baker and Fairlie in terms of Brane-Wave duality based on early work on Universal 
Field Equations [21]) that could be inferred under the exchange: 

$$\partial_{\sigma^a} X^{\mu_1} \leftrightarrow \partial_{\mu_1} \phi^a (x). \eqno (2.22)$$  
this duality (2.22) exchanges target manifold  for base manifold. The composite field strength is then replaced by :

$$[{F}_{\nu_1 \nu_2....\nu_{p+1}} (\phi^a (x)) ]^2 
\leftrightarrow [\epsilon^{\sigma^1 \sigma^2....\sigma^{p+1}}\partial_{\sigma^1}X^{\nu_1}......\partial_{\sigma^{p+1}}X^{\nu_{p+1}}]^2. \eqno (2.23)$$
ths last equation (2.23) has the similar  form as the Skyrmion-based Dolan-Tchakrian conformaly invariant Lagrangians [6]  for the $m$-brane in the case that 
$m+1=2n=2(p+1)$. Similar actions can also be constructed for $m+1=2n+1$ but in such case conformal invariance was lost. A spinning membrane action based on the Dolan-Tchrakian Lagrangians  was constructed in [7]. 
Dolan-Tchrakian have shown that upon the algebraic elimination of the auxiliary world volume metrics $g^{ab}$, via their equations of motion,  one recovers the Nambu-Goto actions. The antisymmetrization w.r.t the indices of Dolan-Tchrakian's  action was done as follows :

$$\int d^{2n}\sigma g^{\sigma^1 \xi^1 }g^{\sigma^2\xi^2 }....g^{\sigma^{p+1}\xi^{p+1} }
\partial_{[\sigma^1}X^{\mu_1}......\partial_{\sigma^{p+1}]}X^{\mu_{p+1}} \partial_{[\xi^1}X_{\mu_1}......\partial_{\xi^{p+1}]}X_{\mu_{p+1}}.\eqno (2.24)$$
the antisymmetrizations of indices $[\sigma^1,\sigma^2....\sigma^{p+1}]$  and 
$[\xi^1,\xi^2....\xi^{p+1}]$
will render such Lagrangians  in the similar form  of eq-(2.23). One can either antisymmetrize w.r.t the world volume metric indices or w.r.t the indices of the derivatives acting on the coordinates $X^\mu$.  Notice, that the latter Skyrmion-based $actions$ are $not$ of the $p+1$-volume squared form ( like the Schild action).The Lagrangian of (2.24) is integrated w.r.t the $2(p+1)$-dim world volume of the $m$-brane, $m+1=2n=2(p+1)$, and $not$ w.r.t the $p+1$-dim world volume of the Schild actions.       

Such actions (2-21, 2-23,2-24) are very suitable to quantize using the Zariski deformation quantization program initiated by Flato et al [16] which has found recent 
interesting applications in the context of $M$(atrix)  theory and deformation of Nambu-Poisson Mechanics [17]. Recently, we were able to show that certain $p$-branes actions could be derived directly from a Moyal deformation quantization of Tchakrian's Generalized Yang Mills Theories [18]. Establishing a relation between branes and 
Generalized Yang Mills should reveal important facts behind Maldacena's $AdS/CFT$ duality conjecture.

\bigskip

\centerline{\bf 3. Towards a New Relativity Principle} 
\bigskip
  
\centerline{\bf 3.1 . The Loop Space Wavefunctional of a $p$-brane : Other approaches } 
\bigskip

A noncovariant Schroedinger-like loop wave equation for the bosonic closed string based on the Eguchi $areal$ time  quantization scheme 
of the Schild action ( which is only area-preserving diffs invariant) was given by [5] :
$$H\Psi [\sigma^{\mu\nu}, A] =-i {\partial \Psi \over \partial A}. ~~~\hbar =1.~~~
\Psi =\Psi e^{-i {\cal E} A}. $$
$$- {1\over l_C} \int_0^1 ds \sqrt {x_\mu '(s)^2 }{1\over 4m^2} ({\delta^2 \over 
\delta \sigma^{\mu\nu} (C) \delta \sigma_{\mu\nu} (C) }) \Psi [\sigma^{\mu\nu}(C)] = {\cal E} \Psi [\sigma^{\mu\nu}(C)]. \eqno (3.1)$$     
the stationary string wave functional, $\Psi [\sigma^{\mu\nu}(C)]$     ,  in this nonrelativistic Schroedinger-like loop wave equation represents the quantum probablity amplitude to find a given closed loop $C$, with spatial area-elements or $holographic$ shadows $\sigma^{\mu\nu}(C)$  onto a spacetime background, as the $only$ boundary of a two-surface of internal area $A$ in a given quantum state $\Psi$. ${\cal E}$ is the energy per unit length or the string tension and $m$ ( discussed earlier)  is the  mass parameter related to the string Regge slope which is not necessary  equal to the variable string tension : ${\cal E}$. 

$l_C$ is the reparametrization invariant length of the loop and the
quantity $ x'_\mu$ represents $d x^\mu /d s$. The operator acting on the $\Psi$ lies inside the integral w.r.t the parameter $s$. This results from the fact that one started from a Hamiltonian $density$ instead of an integrated quantity. The factor ${1\over l_C}$ is due to an average procedure that results from performing the loop derivative by attaching an infinitesimal petal, at any given point along the loop, and then averaging over all points along the loop. The wave equation is explicitly 
$s$-reparametrization invariant despite the fact that the Schild action is only area-preserving diffs invariant. .

To be able to compare (3.1) with the previous equations (2-7, 2-17) it is convenient to multiply (3-1) by $l_Cm^2$ and to reintroduce the ${\cal E}l_C m^2$ term $inside$  the integral as follows : 

$$\int_0^1 ds \sqrt {x_\mu '(s)^2 }{1\over 4} [   {\delta^2 \over 
\delta \sigma^{\mu\nu} (C) \delta \sigma_{\mu\nu} (C) }  + {\cal E}m^2 ] ~\Psi [\sigma^{\mu\nu}(C)] =0 . \eqno (3.2 )$$
Aside from a combinatorial factor of $2$ ( that can be reabsorbed in the tension) the last equation has exactly the same form as (2-7b, 2-18) for the $p=1$ string case. As stated earlier, in the Nambu-Goto string quantization the 
string tension is kept $fixed$ and identified with the $m^2$ parameter : ${\cal E} = T =m^2 ={1\over 2\pi \alpha'} $ . Hence one has an exact result with (2-7b. 2-18) since 
then the second term inside the integral becomes ${\cal E}^2 $.  In this special case one recovers the Marshall-Ramond wave equation obtained long ago [26]. Similar results were obtained by Hosotani [13]. One could reexpress (3.2 ) in terms of ${\delta \over \delta x^\mu (s)}$ derivatives as discussed by [5,13 ] by using the chain rule :

$${\delta \over \delta x^\mu (s)} = {dx^\nu \over ds}{\delta \over \delta \sigma^{\mu\nu} } \Rightarrow 
\int_0^1 ds \sqrt {x_\mu '(s)^2 }~{1\over 4} [   ( {\delta  \over \sqrt { x'_\mu (s)  ^2 } 
\delta x^\mu (s)} )^2  + {\cal E}m^2 ] ~\Psi [x^\mu (s) ] =0 .    ~for~{\cal E}m^2 ={\cal E}^2        \eqno (3.3 )$$
where we have explicitly written (3-3 ) in terms of the $s$-reparametrization invariant expressions :

$${1 \over \sqrt { (x'^\mu)^2 } }{\delta \over \delta x^\mu (s)}. ~~~d\mu (s) = \sqrt { (x'_\mu)^2 } ds. \eqno (3.4)$$

To sum up, we have shown how the covariant Hamilton-Jacobi equation for closed $p$-branes ( $p$-loops) based on Dirac-Nambu-Goto actions yields the correct Schroedinger loop wave equation based on the Eguchi $areal$ quantization of the Schild action upon setting ${\cal E}=m^2$ in the string case. for example. . Inotherwords, the $classical$ field equations for the $p$-branes furnish the first quantized Schroedinger loop wave equations. Naively one would expect that a second quantization will yield a quantum field theory of $p$-branes. Weinberg has strongly advocated the fact that quantum fields are $not~ quantized$ wave functions and that second quantization is a concept that should be banned from Physics [22]. Secondly, there are many authors which are prejudice about the whole field theory approach to strings and $p$-branes [23] . In particular, the fact that string feld theory so far has  been unable to go beyond perturbation theory and to capture $M$ theory is a negative s!
ign that suggets that string field theory is not on the right track. Thirdly, as mentioned earlier, there were $no$ tachyons in the spectrum of [13-15]  and this 
clearly disagrees  with the conventional results obtain from the BRST and BFV methods where tachyons were found in the bosonic string spectrum. Nevertheless we believe that the efforts [13-15] are not in vain.

To resolve these  paradoxes and to tackle these objections   we believe that we must invoke to an $extension$  of the principle of Relativity where all dimensions and signatures must be on the same footing
[8,9].  Hull [24] has recently emphasized this important fact in connection to duality and signature changes in string theory. 
This is the subject of our next section. 

\bigskip
\centerline{\bf 3.2  Hints of a New Relativity Principle : C-Spaces} 
\bigskip

Our  arguments goes as follows [8] : So far we have dealt with the Schroedinger loop wave equation associated with Eguchi $areal$ quantization.   
The $covariant$ Loop wave equation must treat the $areal$ time $A$ on the same footing as the $spatial$ area or holographic components $\sigma^{\mu\nu}(C)$. This led the author [8] to write a Klein-Gordon like Master field equation in a generalized {\bf C}-space where 
point, loops and surface histories were $all$ treated on a single footing and where the extension of the Special Relativiy principle transformed each object 
into one another; i.e the point, loop and surface histories are  reshuffled into each other under the generalized Lorentz transformations which in {\bf C}-space amount to area-preserving antisymmetric matrix-coordinates transformations. 

This is consistent with the ideas of Poly-Dimensional-Relativity of [9] that purports to treat all dimensions and signatures of spacetime into one single framework of multidimensional or Clifford-algebra valued equations . Relativistic transformations in Poly-Dimensional Relativity  transform a {\bf 2}-vector into a {\bf 3}-vector or into a {\bf 5}-vector, for example. This occurs also under paralell transport of a {\bf p}-vector along a closed curve : the rank of the {\bf p}-vector can change. Dimensions/signatures are on the eye of the beholder. It has been argued that duality in 
Electromagnetism is linked to signature change rotations [9] since Clifford algebras do distinguish spacetime signatures [9]. For recent results on string dualities and 
signature changes see Hull [26].       

The covariant loop wave functional equation for histories taken place in $D=4$ required the use of an $eleven$ dimensional  {\bf C}-space of $8\times 8$ antisymmetric matrices, $X^{MN}$ ( whose entries have  area-dimensions ) encoding the point, loop and surface histories [8] . The entries of such matrix have one scalar part representing the proper 
$area$  interval spanned by the motion of the $loop$ in spacetime. Another entries correspond to the $6$ holographic components $\sigma^{\mu\nu} (C)$. 
Another entry corresponds to the $4$ center of mass coordinates of the loop $x^\mu_{CM}$. The total number of degress of freedom in $D=4$ is equal to $1+6+4= 11$. 
The wave equation we wrote is : 

$$ \int d\Sigma ~  [ {\delta^2 \over \delta X^{MN}(\Sigma) \delta X_{MN}(\Sigma)  } +T^2 ]~ \Psi [X^{MN} (\Sigma)] =0. \eqno (3-5)$$
where $\Sigma$ is  an invariant evolution  $parameter$  of $areal$ dimensions extending the role that the proper time of a point particle path had in Special Relativity : 

$$(d\Sigma)^2 ={\cal G}_{MNPG} ~dX^{MN} dX^{PG}.\eqno (3-6)$$
Where    ${\cal G}_{MNPG}$ is a hyper-matrix that generalizes the notion of the ordinary metric tensor $g_{\mu\nu}$ in Special Relativity. The generalized "Lorentz" 
transformations in {\bf C}-space amount to a change of ( antisymmetric) matrix coordinates :

$$X'^{MN} =\Lambda^{MN}_{PQ} (\beta^i,\theta^j,....)  X^{PG}. \eqno (3-7)$$
that ensure that the $areal$  interval is invariant : $(d\Sigma)^2 =(d\Sigma ')^2$. It is in  this strict sense that " Lorentz" transformations in {\bf C}-space  are 
equivalent to $area$ preserving diffs !. Because of the matrix nature of the coordinates, {\bf C}-space is intrinsically Noncommutative from the very start. 
Special Relativity required the introduction of the speed of light, $c$, to embrace space with time. The new principle of 
Relativity which embraces points, loops and areas on the same footing requires the introduction of a $length$ scale : the Planck length $l_P$ [8]. 
If one were to set the $l_P$ to $zero$, after suitable scalings of the matrix entries , the invariant interval (3-6) will reduce to the standard proper time 
interval spanned by a point particle in terms of ordinary 
coordinates. One recovers in that limit ordinary Minkowski spacetime Physics [8]. The extended objects collapse to a point. 
We strongly believe that the 
following arguments below support the crucial role that a length scale has in the New Relativity principle that seems to lurk  behind $p$-brane Quantum Mechanics. 

The Planck length $l_P$ is $not$ the same as the string length $l_s$. It is known that $D$-branes can probe distances smaller than $l_s$. 
In the same fashion that $c$ is the maximum attainable velocity in ( motion) Special Relativity, 
$l_P$ is taken to be the minimum resolution scale in Nature. Nottale [27]  has developed his theory of $Scale$ Relativity by abandoning the naive idea 
that spacetime at the fundamental level is differentiable : it is intrinsically Fractal. The spacetime Riemannian continuum, as we know,  is large distance effect. At Planck 
scales it should be replaced by a transfinite Cantorian-Fractal space  [28].    
The Planck scale is postulated as the minimum
$resolution$  scale in Nature that a physical apparatus can resolve. This doesn't mean that one cannot  have a zero length in Nature : one cannot have zero length with 
$zero$ resolution. Inotherwords, there is no such a thing as a $point$ with absolute certitude in Nature. There is a "point"  up to plus or minus the Planck scale resolution. 
The point is smeared, into a fuzzy "ball" 
and therefore the concept of "dimensions" is lost in the process. Dimensions become resolution dependent. This picture is not 
farfetched, Connes Noncommutative Geometry also abandons the idea of " points". One replaces the ordinary commutative algebra of point-valued coordinates with a noncommutative algebra of matrices, for example.  

Resolutions are $not$ to be confused with Statistical Uncertainities . Based on this principle,  Generalized Heisenberg Relations taken into account the mimimum 
length principle were given in [29].   
In Scale Relativity the Compton wavelength of a particle ceases to be the inverse of the physical momentum. 
It takes an $infinite$  energy to probe the Planck minimal scale resolution.  Likewise, it takes an infinite energy to accelerate a nonzero mass object to the speed of light. 
For this reason we proposed that strings cannot probe distances smaller than the Planck scale [29]. Spacetime at the Planck scale ( spacetime foam) 
is inherently $fractal$ [5]  . This is another reason we why believe Scale Relativity could very well operate in string theory .   
$D$-branes, as said earlier,  can probe distances smaller than $l_s$ ( which is not the same as $l_P$) . 

The Scale Relativity principle also applies to the 
large macroscopic scales that permitted Nottale an elegant, viable and plausible resolution of the cosmological constant problem .  
We do not yet know whether Nature follows the Scale Relativity principle or not. 
All we wish to do is to point out the similarities between Nottales ideas and ours in connection with the necessity to introduce a length scale $l_P$ into the 
New Relativity principle . 
It must be introduced if we wish to treat all dimensions on the same footing. Evenfurther, The combination of $c$ and $l_P$ yields the maximum proper 
acceleration in Nature  : 
${c^2\over l_P}$ that requires an extension of the ordinary spacetime Riemannian Geometry to a more fundamental Finslerian  Geometry of the spacetime tangent bundle [30]  
( not just the geometry of spacetime itself) .  
Strings themselves are subject to a maximum value of tidal forces given by the maximum proper acceleration ${c^2 \over l_P}$.  
When one sets $l_P$ to zero ( infinite proper acceleration ) one recovers in that limit ordinary 
Riemannian geometry from the more fundamental Finslerian  geometry [30]. Finsler geometry is then the intermediate step between General Relativity in ordinary spacetime with 
Relativity in {\bf C}-space.

Having taking this speculative detour on the nature of the length scale we continue with more definitions :  
The extended antisymmetric matrices 
$\Lambda^{MN}_{PQ} (\beta^i,\theta^j,....) $, or hyper-matrices, depend explicitly on the 'angles ' , ' boosts'  parameters : $\beta^i, \theta^j,....$ associated with the invariance group in {\bf C}-space. Shortly we will discuss what the symmetry transformations may look like.

In general, for a membrane history we must have a cubic matrix $X^{MNP}(\Sigma_2)$  and a corresponding hyper-metric ${\cal G}_{M_1M_2......M_6}$ 
where $\Sigma_2$ will be the invariant {\bf 3}-volume parameter interval generalizing the $areal$ invariant interval of (3-6). 
It is the higher dimensional version of proper time in Special Relativity. For $p$ branes one must use $hyper$ matrices  of the form 
$X^{M_1M_2....M_{p+1}} (\Sigma_p) $  with the  corresponding hyper-metric 
${\cal G}_{M_1M_2......M_{2 (p+1)}}$  and to replace the string tension squared terms with the $p$-brane tension squared. The wave equation read identical to (3-5) 
after replacing $\Sigma$ with the $\Sigma_p$ invariant parameter interval ( 'proper time') and the matrices for hyper-matrices.  
The results of [8]  also include  $tensionless$ or null strings/branes . The analog of a ``photon'' in {\bf C}-space are the null strings/null $p$-branes [8]. 
One should replace in that  case the $null$ parameter $\Sigma_p$ with another one in the same way that null geodesics in spacetime  requires the introduction af a 
new nonzero affine parameter $\lambda$ along its trajectory. The wave equations are those corresponding to an extension of Electromagnetism in {\bf C}-space.

Now we turn to the discussion of how to find the transformation hyper-matrices appearing in (3-7). This picture is consistent with the Clifford algebraic approach of 
Pezzaglia [9] who has argued that one could write in $D=4$ the " multivector" valued object : :

$${\cal C} =a +a_\mu \gamma^\mu + a_{\mu \nu} \gamma^{[\mu} \gamma^{\nu]} +.....a_5 \gamma^5. \eqno (3-8)$$  
The automorphism group of the Clifford algebra in $D=4$ will $reshuffle$ the $C$-multivector components : a {\bf 2}-vector turns into a {\bf 3}-vector and so forth. Inotherwords, the $rank$ of the tensors change in this Polydimensional Relativistic approach. We believe this should be taken seriously because this is a nice  geometrical picture which is able to interrelate all extended objects ( dimensions or ranks) and to encompass their duality symmetries in one single footing. In addition signature changes are also incorporated in this picture [9]. 

To find out the transformation properties (3-7) it is convenient to study the point particle case. It is well known ( to the experts ) that the conformal group action in 
$R^4$   on the Lorentz vector $X^\mu$  can be represented as a Mobius-like transformation whose parameters 
{\bf a,b,c,d}  are elements of the Vahlen matrices [25] . This is attained by assigning to the vector $X^\mu$ the Clifford-algebra (matrix-valued ) object :  

$$X^\mu \equiv (x_0,x_1,x_2,x_3) \leftrightarrow X=x_o \gamma^0  +x_1\gamma^1 +x_2\gamma^2 +x_3 \gamma^3 . \eqno (3-9)  $$
( quaternionic and octonionic approaches are also possible for higher dimensions)  
which allows to define the conformal transformation like a Mobius like one :

$$X = (aX+b )(cX +d) ^{-1}. \eqno (3-10)$$
The $2\times 2$ complex Vahlen matrices [25] are matrices whose respective entries {\bf a,b,c,d} are Clifford-algebra valued $Cl(3,1)$. 

The idea now  is to assign to the hyper-matrix $X^{M_1,M_2,.....M_{p+1}}$, belonging to the {\bf C}-space,  the Clifford-algebra-valued  " multivector " containing 
respectively : a scalar, an ordinary vector, a {\bf 2}-vector, a {\bf 3}-vector,.....all the components like the ones in eq-(3-8).  
The scalar would correspond to the invariant proper volume spanned by the $p$-loop. The vector to the center of mass coordinates of the $p$-loop. 
The {\bf 2}-vector to the holographic $area$ components $\sigma^{\mu\nu}$ associated with a loop;  the {\bf 3}-vector to the {\bf 3}-volume holographic $\sigma^{\mu\nu\rho}$
components associated with a {\bf 2}-loop....and so forth.
In this fashion one would have a $1-1$ correspondence between each of the components of the hyper-matrices   $X^{M_1,M_2,.....M_{p+1}}$   and the multivector Clifford-algebra valued object ${\cal C}$ given in eq-(3-8). To be more precise, one has a hierarchy of histories or sets :

$$ S_0 \equiv (\tau, x^\mu).~~~S_1 \equiv (A,x^\mu, \sigma^{\mu\nu} (C)). ~~~S_2 \equiv (V,x^\mu, \sigma^{\mu\nu\rho}
(C_2).......\eqno (3.11)$$ 
where for example, $\tau$ is the proper time elapsed 
by a point particle path of coordinates $x^\mu$. $V$ is the proper volume spanned by the motion
of a {\bf 2}-loop ( membrane) 
through spacetime. 
$\sigma^{\mu\nu\rho}(C_2)$ are the holographic projections of the volume enclosed by the {\bf 2}-loop onto the spacetime coordinate planes. 
$x^\mu$ is the center of mass coordinates of the {\bf 2}-loop. And so forth. In this hierarchy there is a $coincidence$ condition : the center of mass coordinates 
$x^\mu$ agrees for all {\bf p}-loops since all {\bf p}-loop histories collapse to a point history in the $l_P=0$ limit. 
The point history is the $base$ point where the hierarchy is built upon. Loop spaces require a base point . 

This implies then that on dimensional grounds : 

$$\tau ={A\over l_P}={V\over l^2_P}=........{\Omega^{p+1} \over l^{p}_P } \eqno (3.12)$$
Thus, the whole hierarchy of point, loops, membrane,....histories is $encoded$  in the entries of the hyper matrix $X^{M_1,M_2,.....M_{p+1}}$. 
The loop history required the ordinary matrix $X^{MN}$ which can be embedded in a cubic matrix, $X^{MNP}$
and the latter can be embedded in a quartic matrix and so forth until we reach $X^{M_1,M_2,.....M_{p+1}}$, for $p+1=D$, when the spcetime dimension is saturated. 

The second step is to find the analog of the Vahlen matrices ( for {\bf C}-space)  which implemented conformal transformations in $R^4$ in an elegant way (3-10).  
To implement the "Lorentz" transformations (3-7) in {\bf C}-space, one could start by  writing  an equation of the form (3-10) where now 
{\bf X}  are  $multivector$ Clifford-valued objects like (3-8)  
( corresponding to the hyper-matrices ) and the {\bf a,b,c,d}  are entries of the extensions ( generalizations)   of the Vahlen matrices in {\bf C}-space. 
The  generalized " Lorentz" transformations will $reshuffle$  the geometry in such a way that a $point$ history will transform into a loop history, for example ;  
a membrane history into a $5$-brane history ....and so forth. Evenfurther, one can also have an object encompassing all the other $p$-branes as building blocks. In this  unified framework all extended objects, $dimensions$ ,  are treated on the $same$ footing. 

Quantization of spacetime takes place $outside$ spacetime : in {\bf C}-space. This is truly as much background independent as it gets. One does $not$  have a quantization process  $in$ spacetime but $of$ spacetime itself, instead ! 
This is attained  by keeping track  ( in {\bf C}-space )    of the $histories$  of the points, loops, areas, membranes,...$p$-branes $excitations ~of$ spacetime. 
Such excitations are $discretely$  quantized in Planck areas, Planck volumes,....units. 
These are the " gravitons" or basic cells of {\bf C}-space. There are also ( null-like)
" photon" or " electromagnetic " excitations in {\bf C}-space
which correspond to the 
$tensionless$ $p$-branes excitations $of$ spacetime. 
In {\bf C}-space, a {\bf C}-event is a $history$ associated with a point, loop, area,....$p$-brane
The collection of 'events' in {\bf C}-space represent the history of all histories. [8].  
This is similar philosophy behind the extensive research  over the years in  the Twistor, Loop Quantum Gravity, Spin Foam models, Spin Networks, Histories, Topos...approaches to Quantum Gravity by Penrose, Isham, Rovelli, Smolin, Ashtekar, Baez , Finkelstein 
and collaborators.    
{\bf C}-space, is a $categorical$ space 
in the true  algebraic sense of the word and the New Relativity Principle is clearly an extension of the ordinary principle of Relativity  : 
Dimensions and Signatures are $relative$, they are in the eye of the beholder [9]. 

We believe this deserves futher study. The algebra behind transformations (3-7) based on extensions of the Vahlen matrices and the conformal group will be the subject of future investigation. In essence the algebra in {\bf C}-space will be to generalize the ordinary Lorentz algebra with generators $L^{\mu\nu}$ and metric $g_{\mu\nu}$ to one 
whose generators contain more indices : one must find a representation in terms of hyper-matrices to compute the commutation relations. 
The Vahlen matrices Clifford-algebraic approach
is more geometrical especially in so far as selecting the conformal group in (3-10). Conformal invariance is a key ingredient of string theory and twistors; it is no 
surprising that it must show up in {\bf C}-space as well. Moreover, there is also room for quaternion and octonions as well in this Clifford-algebraic approach.      
The ordinary Lorentz algebra reads :

$$[ M^{ab}, M^{cd}] = g^{bc}M^{ad}- g^{bd}M^{ac}+ g^{ad}M^{bc }- g^{ac}M^{bd}. \eqno (3.13)$$
We need a generalization of (3-13) in 
 terms of hypermatrices and also multiple commutators. 
Some interesting realizations of the $triple$  Nambu commutator (bracket) $[A,B,C]$, related to volume-preserving diffs,  were given in [17] in terms  of ordinary and 
$cubic$  matrices as well.  
These results could be very helpful in order to find explicit realizations of the " Lorentz" algebra in {\bf C}-space in terms of cubic, quartic,...$hyper$-matrices and 
multiple commutators.

One can use the lessons from 
Connes Noncommutative Geometry which establish a correspondence between spaces and noncommutative unital star algebras. Inner automorphisms of the noncommutative 
algebras correspond to the internal gauge symmetries of the spaces. Outer automorphims of the algebra correspond to diffeomorphims of the spaces, for example
Since the
automorphisms
groups of the Clifford algebra reshuffle the geometry [9] and these precisely 
correspond 
to the " Lorentz" transformations of the corresponding {\bf C}-space hypermatrices we may reinterpret 
all our formalism as a very special example
of a Noncommutative 
Geometry based on Clifford algebras. 
Automorphisms Groups of Noncommutative Vertex Operator algebras in connection to strings compactifications in $T^d$, 
the Monster group , Bocherds ( Generalized Kac Moody ) algebras and the generation of spacetime gauge symmetries of string theory ( dualites,...) were studied by [31].
The uniqueness of the Monster has led [32] to propose that a 
unique theorie of gravity and matter should be based on the Monster Moonshine Module 
constructions. For a detailed introduction to the Monster and the classification of {\bf CFT} theories we refer to [33].

Schroedinger-like loop wave equations for $p$ branes exactly with the same form as (2-7b, 2-17b) were also discussed briefly ( without derivation) in 
[8]. Pavsic [10] has constructed wave equations for ``wiggled'' $p$ branes based on a Fock-Schwinger 
proper time unconstrained formalism at the expense of introducing auxiliary fields ( the wiggles). His construction also relied on the loop space approach method.

\bigskip

\centerline {\bf 4. }

\bigskip
\centerline{\bf 4.1 Dirac-like Quantum Mechanical Wave Equations for $p$-Branes}
\bigskip

A Dirac-square root like procedure for the Hamilton-Jacobi inspired equation 
(2.7b,2-18) yields :

$$\int d^p\mu (s)~ [ i \Gamma^{\mu_1 \mu_2......\mu_{p+1}} (s)  {\delta  \over \delta 
{\Omega }^{\mu_1 \mu_2......\mu_{p+1}}(s)  } + {\cal E}_p   ]~ \Psi [{\Omega }^{\mu_1 \mu_2......\mu_{p+1}}(C_p)]=0. \eqno (4.1) $$
where no $curvature$ terms have been included. Usually when one takes the square of the Dirac equation curvature terms of the type :

$$[\gamma^\mu \nabla_\mu, \gamma^\nu \nabla_\nu ] \sim \sigma^{\mu\nu} R_{\mu \nu}. \eqno (4.2)$$
appear. This results from the fact that a product of two operators $AB$ can always be written as a sum of a commutator $[A,B] $ and anticommutator $\{ A, B\}$.  The 
commutator give rise to the curvature terms and the anticommutator terms to the D'Alambertian for the ordinary point particle case.  
We set the curvature terms to $zero$ to simplify matters; i.e "flat" $p$-dimensional loop space case.  
A Dirac  equation for the string case similar to (4-1) was given by Hosotani [13]. Eq-(4-1) is the $p$-brane Dirac's  equation  extension of Hosotani's equation. 
The highly nontrivial Lorentz and {\bf C,P,T} transformation properties of the $\Psi$ field of (4-1) 
are beyond the scope of this work. Carson, Ho and Hosotani studied the transformation properties for the string wave functional [14] where  $\Psi$ is $not$ a typical Dirac spinor , it is a more complicated object due to the nature of the generalized Dirac matrices.

The anticommutator of the antisymmetric 
$\Gamma^{\mu_1 \mu_2......\mu_{p+1}} $ matrices is proportional to the unit matrix $I$ : 

$$\{\Gamma^{\mu_1 \mu_2......\mu_{p+1}} (s) , \Gamma^{\nu_1 \nu_2......\nu_{p+1}} (\sigma')   \}
=  2 {\cal G}^{ ([\mu_1 \mu_2......\mu_{p+1}][\nu_1 \nu_2......\nu_{p+1}] )}  \delta^{p} (s-s') I. 
\eqno (4.3)$$
with 
$${\cal G}^{( [\mu_1\mu_2......\mu_{p+1}][\nu_1 \nu_2......\nu_{p+1}]) } =
g^{\mu_1 \nu_1}g^{\mu_2 \nu_2}......g^{\mu_{p+1}\nu_{p+1}} + permutations. \eqno (4.4)$$
the r.h.s of (4-4) can also be written as the determinant of the $(p+1)\times (p+1)$ matrix $G^{ij}$ whose elements are $g^{\mu_i \nu_j}$. 
The permutations ensure that the generalized metric is antisymmetric under any exchange of $\mu_1, \mu_2.....\mu_{p+1}$ and $\nu_1, \nu_2.....\nu_{p+1}$ indices, respectively,   and is symmetric under the exchange of the collective group indices $\mu \leftrightarrow \nu$. Notice that the sum of antisymmetrized  products of 
ordinary metrics appears also in the Dolan-Tchrakian action (2.24).

Since we have assumed that there are $no$ curvatures terms and that for all practical purposes the $p$-loop space space is " flat", to simplify matters one can assume that the matrices  
$\Gamma^{\mu_1 \mu_2......\mu_{p+1}}$ do not have an explicit $s$ dependence. In the ordinary $D=4$ curved spacetime scenario one introduces the vielbeins $e^A_\mu$ in order to define the Clifford algebra in the tangent spacetime and to convert tangent space indices into curved-spacetime  ones. 
In the simplest case when the matrices do not depend on 
$s$ the matrices $\Gamma^{\mu_1 \mu_2......\mu_{p+1}}$ can be written as linear combinations of sums of antisymmetrized products of ordinary Dirac-Clifford algeba matrices :

$$\Gamma^{\mu_1 \mu_2......\mu_{p+1}} =\sum_{i=1}^{2^D} C^{\mu_1 \mu_2......\mu_{p+1}} _{(i)} \Gamma^{(i)}. \eqno (4.5)$$
where a $2^D$ dimensional Clifford algebra basis is spanned by the antisymmetrized products of the ordinary $2^{D/2} \times 2^{D/2}$ $\gamma^\nu$  matrices :   

$$ \Gamma^{(i)} = \{ I, \gamma^{\nu_1}, \gamma^{[\nu_1}\gamma^{\nu_2 ]},....
  \gamma^{[\nu_1}\gamma^{\nu_2 }\gamma^{\nu_3}.....\gamma^{\nu_D ]} \}. 
\eqno (4.6)$$
where $\gamma^\mu $ are the standard gamma matrices in $D$ dimensions. 
Eqs-(4-3) will in principle determine  the algebraic equations to solve for the  tensorial coefficients   $ C^{\mu_1 \mu_2......\mu_{p+1}} _{(i)}$  that define the 
$2^{D/2} \times 2^{D/2}$  $\Gamma^{\mu_1 \mu_2......\mu_{p+1}}$ matrices (4-5) based on the fact that 
each metric component in the r.h.s of eq-(4-4) can be replaced by the anticommutator 
$\{\gamma^\mu, \gamma^\nu \} =2g^{\mu \nu} I$. An extension of the Dirac $p$-brane equation [4-1) in {\bf C}-space, in the same lines as eq-(3-5),  
is also possible by using hyper-matrices and extending the generalized Dirac algebra to one in {\bf C}-space and 
parametrizing the dynamics by the invariant $\Sigma_p$ -parameter interval. 
In essence we are proposing an extension of the ordinary QFT in spacetime to a QFT in {\bf C}-space. To go further implies " curving " the {\bf C}-space, 
introducing connections, torsion, curvatures, and the like. For the time being we must concentrate in the " flat" {\bf C}-space case, find the " Lorentz " algebra, realization of its generators in terms of hypermatrices, ....and then see whether a QFT as we know it can be constructed.

\bigskip
\centerline {\bf 4.2  The De Donder-Weyl- Kanatchikov-Navarro Approach}
\bigskip

A finite dimensional formulation of  QFT based on the De Donder-Weyl Hamiltonian formalism of the 1930's has been undertaken in the 1990's by Kanatchikov and Navarro [11, 12 ] where , among other things, the 
quantization rules and equations of motion are covariant. The approach [11] is  based on Lagrangian systems of the general type $L=L(\phi^a, \partial_\mu \phi^a, x^\mu)$ which may include spacetime sources for the fields $\phi^a$ represented by the coordinates $x^\mu$. .    
The generalized covariant Legendre transform is $H=\pi^\mu_a \partial_\mu \phi^a -L$ associated with the $polymomentum$ field variables 
$$\pi^\mu_a ={\partial L\over  \partial (\partial_\mu \phi^a)} . \eqno (4.7)$$  
Extreme care must be taken $not$ to confuse the latter momentum of eq-(4-7) with the polymomentum   discussed earlier (2-5a, 2-5b,....)!. The  Kanatchikov-Navarro  construction replaces the Schroedinger equation of ordinary Quantum Mehanics with one of the covariant Dirac-like form :

$$i\gamma^\mu \partial_\mu \Psi = {\hat H} \Psi. \eqno (4.8)$$
The units are $\hbar =1$ and , again, a suitable mass/length parameter $\kappa$  
should be inroduced in the l.h.s to match dimensions. This will be done below. The field operators have  the following correspondence principle with the classical field observables : 

$${\hat \phi}^a \rightarrow \phi^a.~~~{\hat \pi}^\mu_a \rightarrow -i\gamma^\mu {\partial \over \partial \phi^a}. \eqno (4.9) $$

In the study of $p$-branes we firstly  must be careful with the proper use of indices. We see the $p$-brane as a $p+1$-dimensional classical field theory associated with a set of $D$ scalar fields $X^\mu (\sigma^a) $.   

If one adpots the conventional view in the literature that the polymomentum is the conjugate to the Jacobians (2-3), the Kanatchikov-Navarro covariant finite dimensional approach to the quantum field theory of $p$-branes based on quadratic actions of the momentum ( a la Schild)  yield for wave equation :

$$\kappa \gamma^a \nabla_a \Psi = {\partial^2 \Psi \over \partial 
{\vec X}^{\mu_1.....\mu_{p+1}} \partial {\vec X}_{\mu_1.....\mu_{p+1}}. } \eqno (4.10)$$
where the covariant derivative $\nabla_a$ is given in terms of the spin connection  $ \partial_a +{1\over 2} \omega_a^{AB}[\gamma_A, \gamma_B]$. The tangent space gamma matrices, $\gamma_A$,   and curved space ones, $\gamma_a$,  are related via the inverse many-bein fields  : $\gamma^a =e^a_A \gamma^A$. Also, ordinary derivatives instead of functional derivatives must be used on the wave function $\Psi ({\vec X}^{\mu_1.....\mu_{p+1}}; \sigma^1,....\sigma^{p+1} ) $ representing the probability amplitude to find a $p$-brane in the quantum state $\Psi$ with 
a $p+1$ world volume ( measure) Jacobian  equal to the $numerical$ value of           
${\vec X}^{\mu_1.....\mu_{p+1}}$ at each world volume point $P$ with coordinates ( seen by  an observer living on the world volume of the $p$-brane)  given by $\sigma^1,....\sigma^{p+1}$. One must reiterate that $\Psi$ is $not$ a functional of all the possible $p$-brane's world volume measure field $configurations$  but, instead,  one has a function of the $numbers$ that the measure takes at each given point $P=(\sigma^1,....\sigma^{p+1}) $.       

The classical Nambu-Goto-Dirac string equations of motion have been
given by Kanatchikov [11] in terms of the De Donder-Weyl string
polymomentum variables   $\pi^\mu_a $  . The author [11] constructed a generalized
graded Poisson bracket for forms of arbitary degree using Lie
derivatives and exterior differentiation in the space of vertical and
horizontal forms associated with an extended polymomentum phase
space. The string  Hamiltonian  was equal to the square root of a
determinant involving quartic powers of the momentum. In a suitable
ortonormal gauge this determinant can be rendered to be quadratic in the momentum
variables as in (4-10). Hence, in this gauge the wave equations for the string based on Kanatchikov's De Donder-Weyl quantization approach  are  :

$$\kappa  \gamma^a \nabla_a \Psi (X^\mu)  = ( -i\gamma^a { \partial \over \partial X^\mu }  )^2 \Psi (X^\mu). \eqno (4.11)$$
The wave function $\Psi (X^\mu)$ is in general a Clifford-algebra valued object : a hypercomplex number [11].  There are some similarities with Pezzaglia's 
equations of 
Polydimensional Relativity [9]. 
The orthonormal gauge simplifies the square root terms. In the
ordinary Nambu string action, the ortho-normal gauge will render the
Nambu-Goto action in polynomial form ( the square root terms simplify)
: 
              
$$ {\partial X^\mu \over \partial \tau}{\partial X_\mu \over
\partial \sigma} =0. ~~~ ({\partial X^\mu \over \partial \tau })^2 =
-({\partial X^\mu \over \partial \sigma})^2 =1. \eqno (4.12)$$  

For general $p$-branes this  gauge choice to simplify the square roots will not work because of the intrinsic $nonlinearities$ of the $p$-branes. 
To solve this problem, a  square root procedure a la Dirac given by the Carson, Ho and Hosotani field theory of $p$-branes [13-15] could be very well employed in the r.h.s of eq-(4-8) 
as well. This warrants further investigation although we believe it will not bring something new to the physics of $p$-branes ;  the upside will be that one will have 
wave equations involving ordinary derivatives, instead of functional ones,  and that one would then avoid the nuissances of QFT ( ultraviolet, infrared)  infinities.
Since the results of [13-15] already dealt with the simplification of the functional differential wave equations by reducing the numbers of degrees of freedom to 
furnish ordinary differential equations, we leave the square root procedure of the r.h.s of (4-8) as an interesting academic exercise.       

The Kanatchikov-Navarro approach to the $p$-brane quantum mechanical wave equations based on the De Donder-Weyl Hamiltonian formalism is a sort of a finite dimensional slice of the ( infinite dimensional )  quantum field theory of $p$-branes based on functional differential wave equations discussed above. Although this finite dimensional 
approach is certainly worth pursuing because it may yield a suitable regularization mechanism to extract finite-valued physically  relevant information from the infinities which plague odinary QFT, we don't know if it will solve  the quantization problem of $p$-branes. One is dealing with very complicated theories, the QFT of $p$-branes, 
which we $insit$  must require the extension of the ordinary Relativity principle to a new one : {\bf C}-space Relativity.         
So far we have dealt with $free$ wave equations. Interactions can naturally be included by means of a potential and the field equations will resemble that of a scalar field with a source proportional to the gradient of the potential w.r.t the field variables. A lot remains to be done.

\bigskip

\centerline{ \bf Acknowledgements .}
\smallskip
We thank Igor V. Kanatchikov for many useful conversations  in Trieste, Italy about his work on the De Donder-Weyl Hamiltonian formalism approach to field theories. 
Also to William Pezzaglia, George Chapline and Eduardo Guendelmann  for many friendly discussions in the Bay Area and Chile. We extend our gratitude to Prof. Rafael  Labarca for his   invitation to the University of Santiago de Chile, Chile. To the Cardenas family and friends for their warm hospitality in Chile were this work was initiated  and to 
the large artistic Perelman family in Chile 
for their many attempts to locate me.

\smallskip

\centerline {\bf Referenes} 

1. A. Aurilia, A. Smailagic and E. Spallucci : Phys. Rev. {\bf D 47} (1993) 

2536.

A. Aurilia , E. Spallucci : Class. Quant. Grav. {\bf 10} (1993) 1217. 

2. H. Kastrup : Phys. Reports {\bf 101} (1983) 1.

3. C. Castro : Int. Jour. Mod. Phys. {\bf A 13} (8) (1998) 1263. 

4. E. Guendelmann, E. Nissimov and S. Pacheva : `` Volume preserving Diffs 

versus Local gauge Symmetries `` hep-th/9505128.

5. S. Ansoldi, A. Aurilia and E. Spallucci : `` Loop Quantum Mechanics and 

the Fractal Structure of Quantum Spacetime `` Jour. of Chaos, Solitons

and Fractals {\bf 10} (2-3) March-April (1999) :  special issue on Superstrings, M, F, S ..Theory.

C. Castro, M. S. El Naschie, eds. 

A. Schild : Phys. Rev {\bf D 16} (1977) 1722.

T. Eguchi : Phys. Rev. Lett {\bf 44} (3) (1980) 126.

6. B. Dolan , D. Tchrakian : Phys. Letters { \bf B 198} (1987) 447.

7. C. Castro : `` The Spinning Membrane and Skyrmions Revisited `` 

hep-th/9707023. 

8. S. Ansoldi, C. Castro and E. Spallucci : Class. Quant. Grav {\bf 16} (1999) 1833-1841

C. Castro : `` The Search for the Origins of M Theory....'' hep-th/9809102.

9. W. Pezzaglia : " Dimensionally Democratic Calculus and Principles of Polydimensional Physics " gr-qc/9912025. 

W. Pezzaglia : `` Polydimensional Relativity, a Classical Generalization 

of the Automorphism Invariance Principle'' in Clifford Algebras and 

Applications in Mathematical Physics `` eds. V. Dietritch. K. Habetha and 

G. Jank. Kluwer Academic Publishers 1997. page 305. gr-qc/9608052.

W. Pezzaglia : `` Should Metric Signature Matter in Clifford Algebra

Formulations of Physical Theories `` gr-qc/9704043.   

10. M. Pavsic : `` The Dirac Nambu Goto $p$-branes as Particular Solutions to a 

General Unconstrained Theory `` Ljubljana IJS-TP/96-10 preprint.

11. I. V. Kanatchikov : Rep. Math. Phys. {\bf 41} (1) (1998). hep-th/9709229.  

I. V. Kanatchikov : `` The De Donder-Weyl Theory and a Hypercomplex 

extension of Quantum Mechanics to Field Theory `` to appear in Rep. Math. Phys 

{\bf 42} (1998).

 I. V. Kanatchikov :  `` From the  De Donder-Weyl Hamiltonian Formalism to 

Quantization of Gravity ``

to appear in the Procc. Int Seminar in Mathematics and Cosmology, Potsdam 1998.

eds. M. Rainer and H. Schmidt. World Scientific 1998.

12-M. Navarro : `` Towards a Finite Dimensional Formulation of QFT `` 

quant-ph/9805010.

13. Y. Hosotani : Phys. Rev. Lett {\bf 55} (1985) 1719.
 
14. L. Carson, Y. Hosotani : Phys. Rev {\bf D 37 (6)} (1988) 1492. 

L. Carson, C.H Ho , Y. Hosotani : Phys. Rev {\bf D 37 (6) } (1988) 1519.

15. C.H Ho : Jour. Math. Phys {\bf 30 (9)} (1989) 2168.  

16. G. Dito, M. Flato, D. Sternheimer and L. Takhtajan : Commun. Mtah. Phys. {\bf 183} (1997) 1. 

17. H. Awata, M. Li, D. Minic and T. Yoneya : " Quantization of the Nambu Brackets " hep-th/9906248.

D. Minic : " M Theory and Deformation Quantization " hep-th/9909022.

18. C. Castro : " Branes from Moyal Deformation Quantization of Generalized Yang Mills Theories " 

hep-th/9908115 submitted to the JHEP. . 

19. C. Castro : " $p$-Branes Quantum Mechanical Wave Equations " hep-th/9812189.

20. L. Baker, D. Fairlie : " Brane-Wave Duality " hep-th/9908157. 

21. D. Fairlie, J. Goaverts and A. Morozov : Nucl. Phys. {\bf B 373} (1992) 214.

22. S. Weinberg : " What is Quantum Field Theory and What Did We Think it is " hep-th/9702027

23. J. Polchinski : Superstring Theory , vols. I, II . Cambridge University Press (1998). 

24. C. Hull : " Duality and Strings, Space and Time " hep-th/9911080. 

25. R. Ablamowicz, P. Lonesto, J. Maks : Conference Report on Clifford Algebras and their Applications 

to Math. Physics. Montpellier, France. September 1989.   

Foundations of Phys. {\bf vol 21} (6) (1991) 735.

26. C. Marshall, P. Ramomd : Nucl. Phys {\bf B 85} (1975) 375. 
 
27. L. Nottale : Fractal Spacetime and Microphysics : Towards a Theory of Scale Relativity . World Scientific 1992. 

La Relativite dans Tous Ses Etats . Hachette Literature. Paris. 1998. 

28. M. El Naschie :  Jour. Chaos, Solitons and Fractals {\bf 10 } (2-3) (1999) 567  . 

29. C. Castro : Found. of Physics Letts {\bf 10} (1997) 273. 

Jour. Chaos, Solitons and Fractals {\bf 10 } (2-3) (1999) 295 .

30. H. Brandt : Jour. Chaos, Solitons and Fractals {\bf 10 } (2-3) (1999) 267. 

31. F. Lizzi, R. Szabo : Jour. Chaos, Solitons and Fractals {\bf 10 } (2-3) (1999) 445 .

32. G. Chapline :Jour. Chaos, Solitons and Fractals {\bf 10 } (2-3) (1999) 311.

33. T. Gannon : " Monstrous Moonshine  and the Classification of Conformal Field Theories " math.QA-9906167.

\bye